\documentclass[jgrga, draft]{agutex}
\usepackage{graphicx}

\newcommand{\apss}{   {\rm Astrophys. Spa. Sci.}\ }

\newcommand{\solphys}{{\rm Solar Phys.}\ }

\newcommand{\ssr}{    {\rm Space Sci. Rev.}\ }
\newcommand{\azh}{ {\rm Sov. Astronom. Zh.}\ }
\newcommand{\physrep}{ {\rm Phys. Rep.}\ }

\authorrunninghead{BIAN ET AL.}

\titlerunninghead{Resonance broadening in quasi-linear relaxation}

\authoraddr{Corresponding author: N.H. Bian
School of Physics and Astronomy,
University of Glasgow, G12 8QQ, Scotland, UK
(n.bian@physics.gla.ac.uk)}

\begin{document}

%
%

\title{Resonance broadening due to particle scattering and mode-coupling
in the quasi-linear relaxation of electron beams}
%
%

%
%



 \authors{Nicolas H. Bian\altaffilmark{1}, Eduard P. Kontar\altaffilmark{1}, and Heather Ratcliffe\altaffilmark{1,2}}

\altaffiltext{1}{SUPA, School of Physics and Astronomy, University of Glasgow, G12 8QQ, Scotland, UK}
\altaffiltext{2}{Centre for Fusion, Space and Astrophysics, Department of Physics, University of Warwick, CV4 7AL, UK}




%
%


\begin{abstract}
Of particular interest for radio and hard X-ray diagnostics of accelerated electrons during solar
flares is the understanding of the basic non-linear mechanisms regulating the relaxation
of electron beams propagating in turbulent plasmas.
In this work, it is shown that in addition to scattering of beam electrons, scattering
of the beam-generated Langmuir waves via for instance mode-coupling, can also result
in broadening of the
wave-particle resonance.
We obtain a resonance-broadened version of weak-turbulence theory with mode-coupling to ion-sound modes.
Resonance broadening is presented here as a unified framework which can quantitatively account for the reduction
and possible suppression of the beam instability due to background scattering of the beam electrons
themselves or due to scattering of the beam-generated Langmuir waves in fluctuating plasmas. Resonance
broadening being essentially equivalent to smoothing
of the electron phase-space distribution, it is used to
construct an intuitive physical picture for the stability of inverted populations of fast electrons
that are commonly observed \emph{in-situ}
to propagate in the solar-wind.
\end{abstract}

%
%

%

\begin{article}

%
%

\section{Introduction}
Solar flares are transient bursts
in the solar atmosphere following the release of energy accumulated in
the coronal magnetic field. A major aspect of solar flares is the high acceleration efficiency
of electrons inferred from their hard X-ray (HXR) emission
\citep[see e.g.][for recent reviews]{2011SSRv..159..107H,2011SSRv..159..301K} .
Although flare accelerated electrons are prominently diagnosed
via their HXR emission at the dense footpoints of coronal magnetic loops,
these electrons can also escape into
interplanetary space and be detected \emph{in-situ} near-Earth or remotely
via their radio emission in the solar wind
\citep[e.g.][as a recent review]{2011SSRv..159..421L}.
As a population of escaping electrons
travel further away from the Sun along the interplanetary magnetic field,
faster electrons outpace the slower ones creating
 an electron beam with positive gradient in velocity space.
 When the number of beam electrons
is sufficiently large, the beam becomes unstable to
the generation of Langmuir waves which then convert
to electromagnetic waves \citep{1958AZh....35..694G}. This plasma emission process can
be seen via type III radio bursts. The rapid frequency drift of these bursts
is associated with the beam travelling away from the Sun in a plasma
of decreasing density, see e.g. \cite{1983SoPh...89..403G, 1985srph.book..177M, 1998SoPh..181..363R,2009IAUS..257..305M} for reviews.
A fundamental problem in the theory of type III solar radio emission is therefore
the understanding of the interaction of the exciter with the background plasma
which is not quiescent nor homogeneous.
For instance, turbulent magnetic fluctuations in the solar wind are
known to efficiently scatter non-thermal particles resulting in broad pitch-angle
distribution despite the effect of adiabatic focusing of the field,
 see e.g. \cite{2009ASSL..362.....S} for a review.
These turbulent magnetic fluctuations also result in cross-field transport of flare accelerated electrons
in the corona \citep{2011A&A...535A..18B,2011ApJ...730L..22K} and in the solar wind \citep{2009ASSL..362.....S}.
Beam generated Langmuir waves can also be efficiently scattered by density inhomogeneities.
The effect of turbulent density fluctuations has been widely discussed in the context
of the beam-plasma relaxation. \cite{1967PlPh....9..719V} first described refraction by random
large-scale density inhomogeneities as a diffusive transfer of Langmuir wave energy in
wave-number space. This diffusion equation in wave-number space
was then used to describe the spectral transfer of wave-energy
out of the region of excitation by the beam \citep{1976JPSJ...41.1757N, 1985SoPh...96..181M}
providing a path for the weakening of the beam instability and also for
acceleration of the electrons during beam-plasma relaxation \citep{1979PhFl...22..321E,2012A&A...539A..43K, 2012ApJ...761..176R}.
Beam-generated Langmuir waves are shifted toward higher/lower phase velocities
when propagating into a region of higher/lower plasma density
so that electrons with higher/lower velocities can interact
with these waves\citep{1969JETP...30..759B}.
Mode-coupling of Langmuir waves, for instance to sound waves,
can also play a stabilizing role on the beam plasma system,
\cite{1974ApJ...190..175P,1977PhRvL..39.1276R} have suggested
that the modulational instability can remove beam-generated Langmuir waves
from resonance with the beam by scattering them to different wavenumbers.
The important role of density fluctuations is attested by the clumpy spatial distribution of the
wave field which is measured \emph{in situ} in the solar wind \citep{1977JGR....82..632G,1981ApJ...251..364L}
as such localization is probably associated with
 density cavities \citep[see e.g. ][]{2010JGRA..115.8103Z,2010JGRA..11510103H}.

In this work we consider anew this important problem of
beam relaxation in turbulent plasmas. We
show that resonance broadening provides a unified framework
to account for the effects of turbulent scattering
of the particles and scattering of the waves in the quasilinear
relaxation of electron beams. Resonance broadening was first discussed by \cite{1966PhFl....9.1773D} in the context
of quasilinear theory in order to treat certain features associated
with the high-intensity of the wave electric field, see also \cite{1971PlPh...13..213R}.
Independently of its origin,
resonance broadening filters out the small scale variations
 of the electron phase-space distribution and therefore, it provides
 an intuitive picture for the weakening and possible suppression of the beam instability
 due to particle and wave scattering.

We start by reviewing in Section \ref{sec:ql_diff} major aspects
 of the quasilinear description of relaxation of electron beams. In Sections \ref{sec:broad_wave}
 and 4,
 it is shown that the effect of scattering of the particles and waves
 can be included in the form of broadening of the wave-particle
 resonance involved in the quasilinear diffusion equations.
 In Section \ref{sec:res_broad}, we start from the \citet{1972JETP...35..908Z} equations
  and derive a resonance broadened version of weak-turbulence theory, where
 broadening of the wave-particle resonance occurs as a result of
 non-local three-wave coupling between two Langmuir (L) modes
 and an ion-sound (S) mode with $k_{L},k_{L'}\gg k_{S}$. A summary
 of the results and conclusions are given in Section \ref{sec:sum}.

\section{Quasilinear diffusion equations for resonant wave-particle interactions}\label{sec:ql_diff}
Quasilinear theory is one of the few theories of plasma turbulence
\citep{1969npt..book.....S, 1972mnpt.book.....D, 1980panp.book.....M,1995lnlp.book.....T}.
It is central to the modelling of stochastic
acceleration of particles by waves and plasma turbulence,
for instance during solar flares \citep[see e.g. ][as reviews]{1997JGR...10214631M, 2012SSRv..173..535P, 2012ApJ...754..103B}.
Originally, quasilinear theory was proposed as a mean-field description of Langmuir wave
generation by a population of beam-electrons, also accounting for
the self-consistent evolution of the wave spectrum that these electrons may emit and reabsorb in the course
of the bump-on-tail instability
  \citep{PhysRev.125.804, 1963JETP...16..682V,1964AnPhy..28..478D}. The quasilinear diffusion equations
describing the relaxation of beam electrons are commonly derived from the Vlasov-Maxwell equations,
which for a one dimensional electron plasma read
\begin{equation}\label{vm1}
\left[\frac{\partial }{\partial t}+v\frac{\partial }{\partial x}+\frac{e}{m_{e}}E(x,t)\frac{\partial }{\partial v}\right]f(x,v,t)=0,
\end{equation}
\begin{equation}\label{vm2}
\frac{\partial E(x,t)}{\partial t}=-4\pi en\int dvvf(x,v,t).
\end{equation}
Here, $f(x,v,t)$ is the electron distribution function (normalized such that $\int dvf(x,v,t)=1$), $E(x,t)$ is the electric field, and $e$, $m_e$ are the electron charge and mass
respectively. In this high frequency limit, the ions are forming a static neutralizing background
and, hence, the electric current $j=e\int dvvf(x,v,t)$ is carried entirely by the electrons.

Let us start by considering the problem of the statistical
acceleration or deceleration of electrons in a broad spectrum of randomly
varying longitudinal electric fields \citep{1966PhRv..141..186S}. The Newton equations of
motion for the electrons under the influence of electric fields
are
\begin{equation}\label{eq:dv_dt}
\frac{dv}{dt}=\frac{e}{m_{e}}E(x,t) \,\,\, , \qquad \frac{dx}{dt}=v
\,\,\, .
\end{equation}
The electric field is assumed to be statistically homogeneous and stationary, with the property
$\langle E\rangle\, =0$ and $\langle E^{2}\rangle\, \neq0$. For the stochastic acceleration
problem at hand, the fluctuating electric field is also characterized by a
a correlation length $\lambda$ and a correlation time $\tau$.
The evolution of the particle distribution function may then be given by a Fokker-Planck equation,
\begin{equation}\label{eqdiffuse}
\frac{\partial f(v,t)}{\partial t}= \frac{\partial }{\partial v}
D\frac{\partial f(v,t)}{\partial v} ,
\end{equation}
where $D$ is the coefficient involved in the diffusion
of electrons in velocity space.
Under this diffusion, particles flow down the gradient in velocity space. This tends to flatten
 this gradient of distribution function and, hence, the action of randomly varying
  electric fields leads to statistical acceleration of the particles
 when $\partial f/\partial v<0$, or to their statistical deceleration when $\partial f/\partial v>0$.
 Accordingly, there is damping or growth of electric energy.
The quasilinear approximation is a perturbation scheme based on the
smallness of the so-called Kubo number, which is essentially the normalized amplitude
of the electric field. Such perturbation scheme is used for obtaining
an expression for the diffusion coefficient $D$ entering Eq.(\ref{eqdiffuse}) in
terms of the spectrum of the electric field fluctuations. The self-consistent rate of damping or growth
of the electric field is then determined by applying conservation of total energy, kinetic plus electric
in the system, in the particular case of waves having specific dispersion relation $\omega(k)$
relating the frequency and wave-number involved in the definition of
the Fourier components of the electric field fluctuations. This proceeds as follow.
The Fourier components ${\hat E}_{k,\omega}$ of $E(x,t)$ are defined via
\begin{equation}
E(x,t) = \sum_{k}\sum_{\omega} \, {\hat E}_{k,\omega} \,
e^{i(kx-\omega t)} ,
\end{equation}
with $k=n\delta k$, $\omega=m\delta \omega$, $\delta k=2\pi/L$, and
$\delta \omega=2\pi/T$.  Here, $n$ and $m$ are integers
while $L$ and $T$ are the spatial and
temporal periods of the electric field. Under the \citet{1959AdGeo...6..441C}
independence hypothesis, the diffusion coefficient in velocity space can be expressed as \citep{2002PhR...360....1K}
\begin{equation}\label{bbd}
D = \frac{e^{2}}{m_{e}^{2}} \sum_{k}\sum_{\omega}
\, |{\hat E}_{k,\omega}|^{2} \, \int_0^\infty dt \, \langle e^{ikx-i\omega t}\rangle \,\,\, .
\end{equation}
In the limit of densely packed Fourier harmonic, one is lead to replace the discrete sum
by a continuous one $
\sum_{k,\omega}\rightarrow \int dk \, d\omega/\delta k \,
\delta \omega \,\,\, ,$
and since by definition the electric field spectrum is given by
\begin{equation}
S_{E}(k,\omega)=|\widehat{E}_{k,\omega}|^{2}/\delta k \,
\delta\omega,
\end{equation}
one also obtains that for a continuous spectrum of fluctuations,
\begin{equation}\label{bb}
D =  \frac{e^{2}}{m_{e}^{2}}\int \int dk \, d\omega  \,
S_{E}(k,\omega) \, \int_0^\infty dt \, \langle e^{ikx-i\omega t}\rangle \,\,\, .
\end{equation}
In this expression for the diffusion coefficient in velocity space, the function given by
\begin{equation}\label{resf}
R(\omega-kv,\Delta \omega)\equiv \int_0^\infty dt \langle e^{ikx-i\omega t}\rangle,
\end{equation}
is the resonance function which characterizes the wave-particle interactions. A more precise
meaning of the averaging procedure $\langle...\rangle$ and
of the resonance width $\Delta \omega$ involved in this expression
for the resonance function will be given below.

We notice that the equations of motion
for the particles can also be written in the following non-dimensional form : $d\widetilde{v}/d\widetilde{t}=\epsilon \widetilde{E}$, $d\widetilde{x}/d\widetilde{t}=\widetilde{v}$.
Time and length have been normalized to $\tau$ and $\lambda$ and $E$ has
been normalized to $\sqrt{\langle E^{2}\rangle}$.
 The Kubo number $\epsilon
=e\tau^{2} \sqrt{\langle E^{2}\rangle}/m_{e} \lambda$
represents the normalized amplitude of the electric field, so that when $\epsilon \ll 1$, the resonance function (\ref{resf}) can be
calculated via the unperturbed particle trajectory. This unperturbed trajectory is given by $
x=vt$ and $v=\mbox{const}$ which are the solutions of the equations of motion
for $\epsilon=0$, i.e. in absence of the field.
The small Kubo number limit corresponds to the weak-field
 limit discussed by \citet{1966PhRv..141..186S}.
In absence of background scattering, the $\epsilon=0$ zeroth-order
particle trajectory is deterministic, hence
we also drop the average involved in the resonance function (\ref{resf}).
In this resonant case, the integral given by $\int_0^\infty dt \, e^{i(kvt-\omega t)} \,\,\, $ is not convergent when $t\rightarrow \infty$, so that an
infinitesimal damping factor $\nu=0+$ is added in the
exponential,
\begin{equation}\label{gg}
\int_0^\infty dt \, e^{i(kvt-\omega t)} \, e^{-\nu t}
= \frac{i}{(kv-\omega) +i\nu} \, \buildrel {\nu \rightarrow 0} \over
{\longrightarrow} \, \pi \, \delta (\omega-kv) + i {\cal P}
\left (\frac{1}{\omega-kv} \right ) \,\,\, ,
\end{equation}
where ${\cal P}$ is the principal value. Only the real part contributes to the resonant diffusion
coefficient~(\ref{bb}) which takes the form
\begin{equation}\label{d}
D=\frac{\pi e^{2}}{m_{e}^{2}} \int \int dk \, d\omega  \,
S_{E}(k,\omega) \, \delta (\omega-kv).
\end{equation}
A Fourier component $[\omega, k]$ of the electric field
fluctuations can exchange energy only with particles having velocity $v$
such that the resonance condition
$v=\omega/k$ is fulfilled.

The only linear modes in Equations (\ref{vm1})-(\ref{vm2}) are Langmuir waves.
Therefore, we now specialize to electric field fluctuations which are produced by Langmuir waves
whose dispersion relation is $\omega(k)\approx \omega_{pe}=\sqrt{4\pi n_{e}e^{2}/m_{e}}$,
and write $S_{E}(k,\omega) = S_{E}(k) \, \delta (\omega-\omega_{pe})$.
It is not here important to include the thermal corrections to the dispersion relation
for $k\lambda_{De}\ll 1$.
The spectral energy density of Langmuir waves is given by $
W(k)=2(S_{E}(k)/8\pi)$, where the factor of two accounts for the electric energy and the
 kinetic energy of the thermal electrons participating in the wave
motion. This leads to the following diffusion equation
\begin{equation}\label{ql1}
\frac{\partial f(v,t)}{\partial t}= \frac{4\pi^{2}
e^{2}}{m_{e}^{2}}\frac{\partial }{\partial v}
\frac{W(k=\omega_{pe}/v,t)}{v} \frac{\partial f(v,t)}{\partial v}\,\,\, .
\end{equation}
describing stochastic acceleration of electrons by a spectrum of Langmuir waves.
Let us come back to the case of a single Fourier component in the electric field ${\hat E}_{k,\omega} \,
e^{i(kx-\omega t)}+\mbox{cc}$, where cc denotes complex conjugate,
so that the diffusion equation is
\begin{equation}\label{dwavedis}
\frac{\partial f(v,t)}{\partial t}= \frac{\pi e^{2}}{m_{e}^{2}} \frac{\partial }{\partial v}|{\hat
E}_{k,\omega}|^{2}\, \delta (\omega-kv)\frac{\partial f(v,t)}{\partial v} \,\,\, .
\end{equation}
For such a single Fourier component, the sharp resonance function $\delta (\omega-kv)$
involved in the diffusion equation (\ref{dwavedis}) implies diffusion
of the electron distribution function $f(v)$ over a single point in velocity space where $v=\omega/k$.
This is a singular and ill-defined diffusion process. A main consequence of the
resonance broadening effect
to be discussed below, is that when the resonance function has a finite width $\Delta \omega$, even a
plane wave can now produce diffusion over a broad region in velocity space. This region
in velocity space is centered around $v=\omega/k$ and has an extent $\Delta v=\Delta \omega/k$
where $\Delta \omega$ is the broadening width. However, we know that a plane wave undergoes damping while accelerating
the particles, and therefore, the singular diffusion equation (\ref{dwavedis}) can still be used
to obtain the average power transferred by the wave to the resonant electrons.
The evolution of electron kinetic
energy is given by
\begin{equation}
\frac{dE_{c}}{d t}\equiv\frac{d}{dt}\langle\frac{1}{2}n_{e}m_{e}v^{2}\rangle=\int dv\frac{\partial f}{\partial
t}\frac{1}{2}n_{e}m_{e}v^{2}.
\end{equation}
Using (\ref{dwavedis}) and one integration by part yields
\begin{equation}\label{enh}
\frac{d E_{c}}{d t}=-\frac{\pi n_{e} e^{2}
|{\hat{E}}_{k,\omega}|^{2}\omega}{m_{e}k^{2}}\left.\frac{\partial
f}{\partial v}\right |_{v=\omega/k},
\end{equation}
consistent with electron acceleration when $\partial f/\partial v<0$
or deceleration when $\partial f/\partial v>0$. Moreover, conservation of the energy given by $
d(|{\hat{E}}_{k,\omega}|^{2}/4\pi+E_{c})/dt=0$, yields
 the following equation, $\partial |{\hat{E}}_{k,\omega}|^{2}/\partial t=2\gamma
|{\hat{E}}_{k,\omega}|^{2}$, describing time evolution of a
given Fourier component of the electric field fluctuations. When these
electric field fluctuations are produced by
Langmuir waves with $\omega\approx\omega_{pe}$, the coefficient $\gamma$ corresponding to Landau damping or growth of the
waves is
\begin{equation}
\label{ld}
\gamma= \frac{\pi \omega_{pe}^{3}}{2k^{2}}\left.\frac{\partial
f}{\partial v}\right|_{v=\omega_{pe}/k}.
\end{equation}
Summing over all Fourier components of the wave field, the evolution of the spectral energy
density is also given by
\begin{equation}\label{ql2}
\frac{\partial W(k,t)}{\partial t}=2\gamma W(k,t).
\end{equation}
Equations (\ref{ql1}) and (\ref{ql2}), are the standard
 quasilinear diffusion equations describing the self-consistent evolution
 of a population of electrons and
 a spectrum of Langmuir waves these electrons may generate\citep{PhysRev.125.804, 1963JETP...16..682V,1964AnPhy..28..478D}. The coefficient $\gamma$,
  giving the rate of growth or damping of each spectral component of the Langmuir wave electric field, is an average rate of stimulated emission and absorption of the waves by the particles, a point
which can be made more transparent in the semi-classical derivation of these quasilinear diffusion
equations \citep{PhysRev.125.804,1980panp.book.....M,1995lnlp.book.....T}.
The quasilinear diffusion equations generally include terms describing the spontaneous
emission of waves and the dynamic friction on the particles.

\section{Broadening of the wave-particle resonances by particle and wave scattering}\label{sec:broad_wave}
The basic rates of wave absorption or emission corresponding to particle acceleration
or deceleration are affected by the presence of
background scattering of particles or waves in the medium.
In the following, the effect of scattering is taken into account in the quasilinear diffusion equations
as a broadening of the wave-particle
resonance. Indeed, the resonant nature of the interactions can be affected by any physical mechanism
which is able to destroy coherence, say on a time scale $\Delta \omega$,
resulting in broadening of the resonance function, for instance into a Lorentzian form:
\begin{equation}\label{dwavedis_Lorentzian}
\pi \delta(\omega-kv)\rightarrow \frac{\Delta
\omega}{(\omega-kv)^{2}+\Delta \omega^{2}} \,\,\, .
\end{equation}
Such Lorentzian form can be obtained by keeping $\nu$ finite in
Equation~(\ref{gg}), making it clear the physical interpretation of resonance broadening
due to wave damping,
\begin{equation}
\Delta \omega=\nu,
\end{equation}
i.e., the resonant interaction between an electron and a Langmuir wave is limited by the
life-time of the wave. Formally, resonance broadening consists in the substitution
\begin{equation}
\pi \delta(\omega-kv)\rightarrow R(\omega-kv,\Delta\omega),
\end{equation}
with $R$ not necessarily a Lorentzian, in the quasilinear diffusion equations.
It is here interesting to recall a point made by
\cite{PhysRev.76.1211}
in a related context, which is that line-broadening
bridges the gap between resonant and non-resonant absorption processes as
the level of scattering in the medium increases.
Such distinction between resonant and non-resonant acceleration by waves,
involving resonance-broadening as a transition, is also at the core
of a recent classification scheme recently proposed by \cite{2012ApJ...754..103B} for stochastic acceleration
during flares.

In the following section, we discuss two important sources of
resonance broadening in the beam-plasma system. These are
the scattering of beam electrons and the scattering of beam-generated Langmuir
waves in turbulent plasmas.

\subsection{Resonance broadening due to particle scattering}
One main effect of background scattering of the particles on the wave-particle interactions
can be understood by modelling such scattering as a
additional random perturbation in the original equations of motion :
\begin{equation} 
\frac{dv}{dt}=\frac{e}{m_{e}}E(x,t)\, , \,\, \qquad \frac{dx}{dt}=v
+\zeta(t)\, ,
\end{equation}
where $\zeta(t)$ is a Gaussian white noise.
Scattering enters here as an external source of stochasticity
in the equations of motion for the particles.
Notice that such scattering does not produce
any change in the kinetic energy of the particles, acting also
in absence of the electric field.
We shall see, however, that scattering
affects the particle acceleration/deceleration rate
due to the
electric field
and, from energy conservation, also
modifies the rate of damping/growth of the electric energy.
In fact, scattering results in broadening of the
wave-particle resonances because it interrupts
at random the coherent emission/absorption of the waves by the particles. In this case,
one has to deal with the random nature of the
integral $
\int_0^\infty dt \langle e^{ikx-i\omega t}\rangle$,
defining the resonance function and, hence, the particle path is written in the form
\begin{equation}
x=vt +\Delta x,
\end{equation}
 where the perturbation $\Delta x$ accounts for the random scattering
of the particles around their free-streaming
trajectory. Under the additional assumption that $\Delta x$ is a Gaussian random process
with $\langle\Delta x^2\rangle\propto t^{\alpha}$, the resonance function becomes broadened
around $\omega-kv$, in the form of a Lorentzian
for $\alpha=1$, of a Gaussian function for $\alpha=2$ or in the form of a Airy function for $\alpha=3$.

 \subsubsection{Spatial diffusion}
 The case $\alpha=1$, with
 \begin{equation}
\langle\Delta x^{2}\rangle=2\kappa t,
 \end{equation}
 represents a standard spatial diffusion of the particles around the free-streaming trajectory, with a spatial
 diffusion coefficient $\kappa$.
For a Gaussian process $\xi$ with zero average, $\langle \xi\rangle=0$, and finite variance $\langle \xi ^2\rangle\neq0$
a cumulant expansion \citep{1962JPSJ...17.1100K} provides the relation
\begin{equation}
\langle e^{\xi}\rangle=e^{{\langle \xi ^{2}\rangle}/{2}},
\end{equation}
where the brackets denote averaging over the probability distribution function (PDF)
of $\xi$, here assumed to be Gaussian. There is a-priori no need to restrict to Gaussian PDFs
except as a simplifying assumption.
Taking $\xi=ik\Delta x$,
we obtain that the resonance function is given by
 \begin{equation}
 R(\omega-kv,\Delta \omega)= \int_{0}^{\infty}dt e^{ikvt-i\omega t}e^{-\Delta\omega t}=\frac{\Delta \omega}{(\omega-kv)^{2}+\Delta \omega^{2}}\, ,
 \end{equation}
 where the resonance width
  $\Delta \omega$ is given by the diffusive time-scale over one wavelength of the electric field, corresponding to
 \begin{equation}
 \Delta \omega=\kappa k^{2}.
 \end{equation}
\subsubsection{Velocity-space diffusion}
Equivalently, the perturbations in the particle path can be written in terms of velocity fluctuations
\begin{equation}
x=(v+\Delta v)t,
\end{equation}
and, furthermore, we assume that the fluctuations
$\Delta v$ evolve diffusively,
\begin{equation}\label{dh}
\langle\Delta v^{2}\rangle=2Dt,
\end{equation}
where $D$ is the diffusion coefficient in velocity space. As a consequence,
the resonance function becomes broadened around $\omega-kv$ as
\begin{equation}\label{pp}
R(\omega-kv,\Delta \omega)=\int_{0}^{\infty}dt e^{ikvt-i\omega t}e^{-(\Delta\omega
t)^{3}}=\frac{\pi}{\Delta\omega}C[i(\omega-kv)/\Delta\omega].
\end{equation}
This expression involves a Airy function $C(z)$ which satisfies the equation $d^{2}C/dz^{2}-zC=\pi^{-1}$ \citep{1980tisp.book.....G}.
The resonance width is given by
\begin{equation}\label{dup}
\Delta\omega=(k^{2}D)^{1/3}.
\end{equation}
Let us provide few examples of possible sources of velocity-space diffusion.
In the original work by \cite{1966PhFl....9.1773D}, the source of velocity-space diffusion is
the wave electric field itself. This means that the resonance function (\ref{pp}) involved in the expression
for the velocity space diffusion $D$ given by Eq.(\ref{bb}) is now a function of $D$, hence we
can rewrite (\ref{bb}) in the form
\begin{equation}\label{bb2}
D =  \frac{e^{2}}{m_{e}^{2}}\int \int dk \, d\omega  \,
S_{E}(k,\omega) \, \int_0^\infty dt \, e^{ikvt-i\omega t} e^{-k^{2}Dt^{3}} \,\,\, .
\end{equation}
This is a non-linear integral equation for $D$. If we take
the limit of weak fluctuations amplitude, $D\rightarrow 0$ in evaluating the resonance function $R$ then
(\ref{bb2}) reduces to the familiar quasilinear result (\ref{d}) with a sharp resonance function
, i.e. $R(\omega-kv,\Delta \omega\rightarrow 0)\rightarrow \pi \delta (\omega-kv)$. In other
words, the quasilinear equations are modified by broadening as the strength of the electric field fluctuations
becomes large. Physically, the resonance broadening effect corresponds to
secular diffusion of particle trajectories which results from the interaction of the particles with the turbulent
electric field fluctuations. We notice that a Gaussian velocity space diffusion with
$\langle\Delta v ^{2}\rangle\propto t$ leads to Gaussian spatial super-diffusion with
$\langle\Delta x^{2}\rangle \propto t^{3}$, hence the form of the resonance function (\ref{pp}).

The source of velocity-space diffusion involved in the broadening of the
wave-particle resonance may also be unrelated to the wave electric field but instead
due to background scattering of the particles for instance at constant kinetic energy.
Beam electrons strongly interact with the ambient magnetic field as the latter
imposes a direction to their propagation.
Moreover, magnetic fluctuations around the mean ambient field produce pitch-angle scattering
which acts as an additional source of
velocity space diffusion and, hence, of resonance broadening.
Pitch-angle scattering results in fluctuations in the partition between translational
(along the ambient field) and rotational energy of the particles in a magnetized
plasma and when this is due to a spectrum of low-frequency magnetic perturbations,
pitch-angle scattering occurs at constant kinetic energy $E_{c}$.
A rigourous treatment of pitch-angle scattering would require a 3D analysis, however
let us emphasize few main aspects of it in a 1D model.
Let us still call $v$ the velocity parallel to the magnetic field, $V=\sqrt{2E_{c}/m_{e}}$ the electron speed and $\mu=v/V$ the pitch-angle cosine.
Interactions of the electrons with magnetic field fluctuations
produce increments of pitch-angle cosine $\Delta \mu$ at constant $V$, and hence increments
of parallel velocity $\Delta v=V\Delta \mu$. Squaring the result and averaging leads to
\begin{equation}\label{pan}
\langle\Delta v^{2}\rangle=\frac{2E_{c}}{m_{e}}\langle\Delta \mu^{2}\rangle.
\end{equation}
Moreover, let us assume that pitch-angle scattering of the particles
produces pitch-angle diffusion with
\begin{equation}
\langle \Delta \mu ^{2} \rangle =2D^{(\mu)}t,
\end{equation}
where $D^{(\mu)}$ is the pitch-angle diffusion coefficient. Hence, pitch angle
scattering acts as a source of diffusion in velocity space but at constant kinetic
energy $E_{c}$, i.e.
\begin{equation}\label{pan2}
\langle \Delta v^{2} \rangle =2\frac{2E_{c}D^{(\mu)}}{m_{e}}t
\end{equation}
a relation which is analogous to (\ref{dh}).
And therefore, the resonance function becomes broadened with a width
given by (\ref{dup}) where $D$ has now the meaning of the coefficient of diffusion for the parallel
velocity at constant energy, which according to $(\ref{pan2})$ tantamount to replacing $D\rightarrow(2E_{c}/m_{e})D^{(\mu)}$
in (\ref{dup}), leading to
\begin{equation}\label{bpit}
\Delta \omega=\left(\frac{2E_{c}D^{(\mu)}k^{2}}{m_{e}}\right)^{1/3},
\end{equation}
which is the resonance width associated with
such scattering mechanism. The above reasoning is due to
 \cite{1973Ap&SS..25..471V} in his treatment of resonance broadening by pitch-angle scattering.
Resonance broadening due to collisions, $D^{(\mu)}$ taking its collisional value
value in (\ref{bpit}) is discussed by \citet{1992wapl.book.....S}.

\subsection{Resonance broadening due to wave scattering}

While resonance broadening due to scattering of particles has been widely discussed
in the past, there is also a possible source of resonance broadening, which is not due to particle scattering
but which is instead due to wave scattering.

\subsubsection{Wave-number diffusion}

For clarity, we first discuss the effect of wave scattering through the following model, which is a variant
 of the Kubo-Anderson
oscillator \citep{1954JPSJ....9..316A,1954JPSJ....9..935K}. The model describes the motion of electrons,
and their stochastic acceleration or deceleration, in a plane wave electric field
whose wave-number is a random function of time:
\begin{equation}\label{s1}
\frac{dv}{dt}=\frac{e}{m_{e}}\left({\hat E}_{k,\omega} \, e^{i[(k+\Delta
k)x-\omega t]}+\mbox{cc}\right)\, , \,\, \qquad \frac{dx}{dt}=v, \,\,\, \qquad
\frac{d\Delta k}{dt}=\zeta(t),
\end{equation}
where, for simplicity, $\zeta(t)$ is taken to be a Gaussian white noise.
According to (\ref{bb}) the expression for the diffusion coefficient in velocity space is given by
\begin{equation}
D= \frac{e^{2}}{m_{e}^{2}}|{\hat E}_{k,\omega}|^{2} \,
\int_0^\infty dt \, e^{ikvt-i\omega t}\langle e^{i\Delta kvt}\rangle \,\,\, .
\end{equation}
The evolution of the wave-number in this model is
a standard diffusion :
\begin{equation}
\langle\Delta k^{2}\rangle=2D^{*}t,
\end{equation}
with $D^{*}$ the diffusion coefficient in wave-number space.
Therefore, we obtain that the diffusion coefficient in velocity space $D$
is related to the diffusion coefficient in wave-number space $D^{*}$ by
\begin{equation}
D= \frac{e^{2}}{m_{e}^{2}}|{\hat E}_{k,\omega}|^{2} \,
\int_0^\infty dt \, e^{ikvt-i\omega t}e^{-D^{*}v^{2}t^{3}} \,\,\, .
\end{equation}
We have implicitly assumed that the effect of
particles scattering is negligibly small
compared to that of the waves.
Equivalently, the diffusion coefficient in velocity space can be written in the form
\begin{equation}
D= \frac{e^{2}}{m_{e}^{2}} |{\hat E}_{k,\omega}|^{2} \,
R(\omega-kv,\Delta \omega) \,\,\, ,
\end{equation}
with the resonance function still involving the Airy function given by (\ref{pp})
but the resonance width is now given by
\begin{equation}
\Delta\omega=(D^{*}v^{2})^{1/3},
\end{equation}
with $D^{*}$ the diffusion coefficient in $k$-space.
While the plane wave is randomly scattered it also undergoes Landau damping when
accelerating the particles. The
average power transferred to the particles by a wave undergoing random change in $k$ leads to
the coefficient of Landau damping of this wave which can be calculated in a manner
similar to what was done for obtaining the Landau damping rate (\ref{ld})
of an unscattered wave. The result is
\begin{equation}\label{ld2}
\gamma= \frac{\omega_{pe}^{3}}{2k^{2}} \int dv
k\frac{\partial f}{\partial v}R(\omega-kv,\Delta
\omega),
\end{equation}
which reduces to the standard expression for Landau damping
of a Langmuir wave given by (\ref{ld}) when $\omega\approx \omega_{pe}$ and when
the level of wave scattering becomes small, i.e. $D^{*}\rightarrow 0$ or $\Delta \omega\rightarrow 0$.

Wave-number diffusion has been widely discussed in the context of the modelling of the propagation
of Langmuir waves in a plasma with background density fluctuations. Above, we have shown that such wave-number
diffusion leads to
broadening of the wave-particle resonances.
In the following, we derive a system
of resonance-broadened quasilinear diffusion equations which describes
the relaxation of an electron beam in a plasma with ambient density fluctuations.

\subsubsection{Broadening of the wave-particle resonances in a plasma with density fluctuations}
The dynamics of both electrons and Langmuir waves
are still taken along $x$, the direction of the external magnetic field.
The plasma density is written as
$n_e[1+\tilde{n}(x, t)]$ with $n_e$ the constant background density
and $\tilde{n}(x,t)$ the relative density fluctuation, which is
assumed to be weak, i.e. $\tilde{n}(x,t)\ll1$. In general, the
characteristic wavenumber $q$ of low-frequency density fluctuations
is much smaller than the characteristic wavenumber $k$ associated
with high-frequency Langmuir waves, so we can make the WKB
approximation and treat the Langmuir waves as plasmons or quasi-particles.
Ambient density fluctuations are associated with a change in the
local refractive index experienced by the waves, so that
wave scattering can be modeled by the equations of motion for the
Langmuir plasmons,
which are given by
\begin{equation}\label{eqn:WKB1}
\frac{d k}{d t} =-\frac{1}{2}
\omega_{pe} \frac{\partial\tilde{n}}{\partial x}\equiv F(x,t),
\end{equation}
\begin{equation}
\frac {d x}{d t} ={v}_g,
\end{equation}
where $F(x,t)$ is the "refraction force" acting on the
quasi-particles, $\omega_{pe}=\sqrt{4\pi n_e e^2/m_e}$ is the local plasma frequency
and ${v}_g=3{v}_{Te}^2 k/\omega_{pe}$ is the group velocity of the quasi-particles.
From Equation (\ref{eqn:WKB1}), we see that random refraction induces stochastic
change in wavenumber, resulting in a diffusion of the wave energy
density in $k$-space. An expression for the diffusion coefficient of wave energy
in $k$-space is thus derived, in the very same quasilinear approximation,
\begin{equation}\label{eqn:diff_1d}
D^{*}= \frac{\omega_{pe}^2
\pi^2}{2} \int_{-\infty}^\infty dq \int_{-\infty}^\infty d\Omega\;
q^2 S_n(q, \Omega)\delta(\Omega-q{v}_g),
\end{equation}
where by definition $\langle\tilde{n}^2\rangle=
\int_{-\infty}^\infty dq \int_{-\infty}^\infty d\Omega\;
S_n(q,\Omega)$. In the particular case where density fluctuations
are due to low-frequency compressive waves with a specific dispersion
relation $\Omega=\Omega(q)$, then
$S_n(q,\Omega)=S_n(q)\delta(\Omega-\Omega(q))$ and $ D^{*}=
(\omega_{pe}^2 \pi^2/2)\int_{-\infty}^\infty dq q^2
S_n(q)\delta(\Omega(q)-q{v}_g)$. A set of
quasilinear equations describing the interaction between Langmuir
waves and beam electrons in the presence of background density
fluctuations is therefore,
\begin{equation}\label{eqdiffuse1}
\frac{\partial f(v,t)}{\partial t}= \frac{4\pi^{2}
e^{2}}{m_{e}^{2}}\frac{\partial }{\partial v}
\frac{W(k=\omega_{pe}/v,t)}{v} \frac{\partial f(v,t)}{\partial v}
\,\,\, ,
\end{equation}
\begin{equation}
\frac{\partial W(k,t)}{\partial t}= \frac{\pi\omega^{3}_{pe}}{k^{2}} W(k,t) \left. \frac{\partial f(v,t)}
{\partial v}\right|_{v=\omega_{pe}/k}+\frac{\partial }{\partial k}
D^{*}\frac{\partial W(k,t)}{\partial k}\,\,\, ,
\end{equation}
which includes a wave-number diffusion in the kinetic equation for
the plasmons.
As a consequence, this system is equivalent to the following quasilinear
diffusion equations with a broadened resonance function :
\begin{equation}\label{eqdiffuse2}
\frac{\partial f(v,t)}{\partial t}= \frac{4\pi
e^{2}}{m_{e}^{2}}\frac{\partial }{\partial v} \int dk
W(k,t)R(\omega_{pe}-kv,\Delta \omega) \frac{\partial
f(v,t)}{\partial v} \,\,\, ,
\end{equation}
\begin{equation}\label{eqdiffuse3}
\frac{\partial W(k,t)}{\partial t}= \frac{\omega_{pe}^{3}}{k^{2}} \int dv
k\frac{\partial f(v,t)}{\partial v}R(\omega_{pe}-kv,\Delta
\omega)W(k,t),
\end{equation}
where the resonance function is given by (\ref{pp}) and the resonance width is
\begin{equation}\label{eqdiffuse4}
\Delta\omega=(D^{*}v^{2})^{1/3},
\end{equation}
with $D^{*}$ given by (\ref{eqn:diff_1d}).

\section{Generalization to three dimensions}

Let us consider the generalization of the quasilinear diffusion equations
to three dimensions including a broadened resonance function. These are
\begin{equation}\label{dwavedis2}
\frac{\partial f(\mathbf{v},t)}{\partial t}= \frac{4\pi e^{2}}{m_{e}^{2}} \frac{\partial }{\partial v_{i}}\int d\mathbf{k} \frac{k_{i}k_{j}}{k^{2}}W_{\mathbf{k}}\, R (\omega_{pe}-\mathbf{k}.\mathbf{v},\Delta\omega)\frac{\partial f(\mathbf{v},t)}{\partial v_{j}} \,\,\, ,
\end{equation}
describing the diffusion of resonant electrons in velocity space, and
\begin{equation}\label{ql2bis}
\frac{\partial W_{\mathbf{k}}}{\partial t}= \omega_{pe}^{3}W_{\mathbf{k}}\int d\mathbf{v}\frac{\mathbf{k}}{k^{2}}
.\frac{\partial f(\mathbf{v},t)}{\partial \mathbf{v}}R(\omega_{pe}-\mathbf{k}.\mathbf{v},\Delta \omega),
\end{equation}
giving the rate of growth or damping of spectral energy in the Langmuir waves. They reduce
to the standard 3D quasilinear diffusion equations when
the broadening width goes to zero, i.e. $R(\omega_{pe}-\mathbf{k}.\mathbf{v},\Delta \omega\rightarrow 0)
\rightarrow \pi \delta(\omega_{pe}-\mathbf{k}.\mathbf{v})$.
The resonance function appearing in these equations is defined as
\begin{equation}
R(\omega_{pe}-\mathbf{k}.\mathbf{v},\Delta\omega)=\int dt \langle e^{i\mathbf {k}.\mathbf{v}t-i\omega_{pe}t}\rangle.
\end{equation}
Let us first consider the effect of random
perturbations $\Delta \mathbf{v}$ in the electron velocity,
i.e. $\mathbf{v}\rightarrow \mathbf{v}+\Delta\mathbf{v}$, hence
\begin{equation}\label{rwi}
R(\omega_{pe}-\mathbf{k}.\mathbf{v},\Delta\omega)=
\int dt e^{i\mathbf{k}.\mathbf{v}t-i\omega_{pe}t}\langle e^{i\mathbf{k}.\Delta \mathbf{v}t}\rangle=
\int dt e^{i\mathbf {k}.\mathbf{v}t-i\omega_{pe}t}e^{-\langle(\mathbf{k}.\Delta \mathbf{v})^{2}\rangle t^{2}/2}.
\end{equation}
Introducing the diffusion tensor in velocity space, which is defined via
\begin{equation}
\langle\Delta v_{i}\Delta v_{j}\rangle=2D_{ij}t,
\end{equation}
we obtain from (\ref{rwi}) the following expression for the resonance function
\begin{equation}\label{rfy}
R(\omega_{pe}-\mathbf{k}.\mathbf{v},\Delta\omega)=\int dt e^{i\mathbf {k}.\mathbf{v}t-i\omega_{pe}t}e^{-(\Delta \omega t)^{3}},
\end{equation}
where the resonance width $\Delta \omega$ is given by
\begin{equation}
\Delta \omega=D_{ij}k_{i}k_{j}.
\end{equation}
Summation over repeated indices is implicitly assumed in the previous expression.
In the same way, let us consider random perturbations $\Delta \mathbf{k}$ in the wave-vector,
i.e. $\mathbf{k}\rightarrow\mathbf{k}+ \Delta \mathbf{k}$, hence
\begin{equation}
R(\omega_{pe}-\mathbf{k}.\mathbf{v},\Delta\omega)=
\int dt e^{i\mathbf{k}.\mathbf{v}t-i\omega_{pe}t}\langle e^{i\Delta \mathbf{k}.\mathbf{v}t}\rangle=
\int dt e^{i\mathbf {k}.\mathbf{v}t-i\omega_{pe}t}e^{-\langle(\Delta\mathbf{k}.\mathbf{v})^{2}\rangle t^{2}/2}
\end{equation}
Introducing the diffusion tensor in wave-vector space,
\begin{equation}
\langle \Delta k_{i}\Delta k_{j}\rangle=2D^{*}_{ij}t,
\end{equation}
we can again write the resonance function in the form of a Airy
function given by Eq.(\ref{rfy}) but
with the resonance width $\Delta \omega$ given by
\begin{equation}\label{wyu}
\Delta \omega=D^{*}_{ij}v_{i}v_{j}.
\end{equation}
Broadening of the wave-particle resonance occurs here as a result of the random refraction of beam-generated
Langmuir waves as they propagate in fluctuating three-dimensional density fluctuations, and therefore,
\begin{equation}
\frac{d \mathbf{k}}{d t} =-\frac{1}{2}
\omega_{pe} \nabla\tilde{n},
\end{equation}
\begin{equation}
\frac {d \mathbf{x}}{d t} =\mathbf{v}_{g},
\end{equation}
where $\tilde{n}=n/n_{0}$ is the relative level of density fluctuations and $\mathbf{v}_{g}=3\lambda_{De}^{2}\omega_{pe}\mathbf{k}$ is the group velocity.
The resonance width (\ref{wyu}) involves the diffusion tensor $D^{*}_{ij}$ in wave-number space, which is given
by
\begin{equation}\label{oip}
D^{*}_{ij}=\frac{\pi \omega_{pe}^{2}}{4}\int d\mathbf{q}\, \int d\Omega q_{i}q_{j}S_{n}(\mathbf{q},\Omega)\delta (\Omega-\mathbf{q}.\mathbf{v}_{g}),
\end{equation}
where by definition $\langle\tilde{n}^2\rangle=
\int_{-\infty}^\infty d\mathbf{q} \int_{-\infty}^\infty d\Omega\;
S_n(\mathbf{q},\Omega)$, meaning that $S_n(\mathbf{q},\Omega)$
is the spectrum of (relative) density fluctuations.
We here assume a given level of large-scale density fluctuations of an
unspecified origin. The density fluctuations
do not have to be produced by waves in which case $\Omega$ and $\mathbf{q}$ are
not related by a specific dispersion relation $\Omega(\mathbf{q})$.
However, when the background density
fluctuations are produced by low-frequency compressive waves,
self-consistency dictates that energy change in the Langmuir
quasiparticles is accompanied by energy change in the
low-frequency compressive waves resulting in Landau damping/growth
of the low-frequency modes on the high-frequency modes, the latter behaving as
quasi-particles \citep{1967PlPh....9..719V}. In order to
respect energy conservation, the system must be closed by a third equation
for the level of compressive density fluctuations which results in $D_{ij}^{*}$ becoming a dynamical variable,
and, hence, also the resonance width $\Delta \omega$, with their time-dependence adjusting self-consistently
to the non-linear channeling of the energy between the
various modes of motion.
In other words, the resonance broadening effect due to Langmuir wave scattering by low-frequency compressive waves
is essentially a manifestation of mode coupling. In the following we focus on mode-coupling to ion sound waves
but Langmuir wave scattering may be due to other compressive modes, such as kinetic Alfven waves
\citep{2010A&A...519A.114B,2010PhPl...17f2308B}. The spectral properties of these electromagnetic compressive
modes are well-documented in the solar wind \citep{2005PhRvL..94u5002B,2013ApJ...768L..10M}.
\subsection{Angular scattering only}

Before continuing to the topic of resonance broadening by mode-coupling, let us discuss briefly the
problem involved, in general, in determining the shape of the resonance function and,
 in particular, in the case where there is only angular scattering of the particles or only angular scattering of the waves. The resonance function
entering the quasilinear diffusion equations
involves the quantity $
\omega_{pe}-\mathbf{k}.\mathbf{v}=
\omega_{pe}-kv[\sin\theta\sin\theta'(\cos\phi\cos\phi'+\sin\phi\sin\phi')+\cos\theta\cos\theta']$,
which is written here in the spherical coordinate systems $(v,\theta,\phi)$ and $(k,\theta',\phi')$ in velocity space
and wave-number space, respectively. An essential problem arises when trying to determine the
shape of the resonance function
when diffusion occurs only in angle, $\theta$ for the particles or $\theta'$ for the waves. The problem is
that none of these diffusion processes are characterized by Gaussian PDFs. This is quite obvious by noticing
that these PDfs must be periodic functions of the angles involved and, therefore, cannot be Gaussian.
There are four angles involved in the problem and they
 can be reduced to two by restricting our considerations
to not too large angular spread of the particles or the waves around $\theta \sim 0$ or $\theta'\sim 0$
corresponding to the direction of the mean field, i.e.
\begin{equation}
\omega_{pe}-\mathbf{k}.\mathbf{v}=\omega_{pe}-kv\cos\theta\cos\theta'=
\omega_{pe}-kv(1-\frac{\theta^{2}}{2})(1-\frac{\theta'^{2}}{2}).
\end{equation}
Let us first consider angular scattering
of the particles assuming a narrow distribution of waves having $\theta'=0$, such that
\begin{equation}
\omega_{pe}-\mathbf{k}.\mathbf{v}=\omega_{pe}-kv(1-\frac{\theta^{2}}{2}).
\end{equation}
Pitch-angle diffusion of the particles results in that $P(\theta,t)$, the PDF
for the pitch-angle $\theta$ evolves according to the diffusion equation
\begin{equation}
\frac{\partial P(\theta,t)}{\partial t}=\frac{1}{\theta}\frac{\partial}{\partial \theta} \theta D_{\theta\theta}\frac
{\partial P(\theta,t)}{\partial \theta},
\end{equation}
when $\sin \theta \sim \theta$. Here, $D_{\theta\theta}$ is the component of the diffusion tensor
corresponding to particle scattering in $\theta$ only. The solution of this equation
can be approximated by a Gaussian only for $\theta \sim 0$ and when $D_{\theta\theta}t\ll 1$,
in which case
\begin{equation}
<\theta^{2}>=2D_{\theta\theta}t.
\end{equation}
This the domain of validity of the approximate approach discussed in Section III above, for the treatment in 1D of
resonance broadening by pitch-angle scattering of the particles,
where we had introduced instead of $\theta$ the notation $\mu=\cos\theta$ for the pitch-angle cosine
of the particles. Outside of this domain of validity, analytical determination of the resonance function must rely
on the more accurate PDF $P(\mu,t)$ which is known to be given by series expansion in terms of Legendre polynomials.
The averaging over $P(\mu,t)$ which is involved in the determination of the resonance
function becomes a more complicated analytical procedure, probably amenable to numerical solution only. However,
the above approximate treatment provided us with an estimate for the resonance width, despite that the precise "line-shape"
may not well be the correct one.

The perfectly
symmetric situation of angular scattering of the waves at constant modulus $k$
requires similar considerations, as in this case
\begin{equation}
\omega_{pe}-\mathbf{k}.\mathbf{v}=\omega_{pe}-kv(1-\frac{\theta'^{2}}{2}),
\end{equation}
and angular diffusion of the waves results in that $P(\theta',t)$, the PDF
for the angle $\theta'$ evolves according to the diffusion relation
\begin{equation}
<\theta^{2}>=2D^{*}_{\theta'\theta'}t,
\end{equation}
where $D^{*}_{\theta'\theta'}$ is the component of the diffusion tensor
corresponding to wave scattering in $\theta'$ only.
Treatment of the simultaneous scattering of waves and particles is also
not quite a simple task as this requires determination of the joint PDF $P(\theta,\theta',t)$
in order to carry the averaging procedure with respects to these random variables.

\section{Resonance broadening due to mode-coupling of Lamgmuir waves with ion sound waves
and weak-turbulence theory}\label{sec:res_broad}

Let us focus on the specific situation of
coupling to the ion-sound waves described by the \citet{1972JETP...35..908Z} equations
\begin{equation}\label{z1}
\frac{i}{\omega_{pe}}\frac{\partial E}{\partial t}+\frac{3}{2}\lambda_{De}^{2}\nabla^{2}E=\left(\frac{n}{2n_{0}}\right)E,
 \end{equation}
\begin{equation}
n_{0}m_{i}\left[\frac{\partial ^{2}}{\partial t^{2}}-c_{s}^2\nabla^{2}\right]\left(\frac{n}{n_{0}}\right)=\frac{1}{4\pi}\nabla^{2}|E|^{2},
\end{equation}
where $c_s$ is the sound speed and $\lambda_{De}$ is the Debye length.
The first equation is a Schrodinger equation describing the evolution of the Langmuir
wave electric field in low-frequency density inhomogeneities.
The second equation describes the dynamical evolution of these density inhomogeneities
when the latter are assumed to be produced by sound waves.
These wave equations are coupled through the terms on their right-hand side.
These terms describe refraction and ponderomotive effects, e.g. \citep{1978PhR....43...43T,
1997RvMP...69..507R}. The Zakharov equations exploit the
separation of time scales between the dynamics of Langmuir and ion-sound waves, so that
the ponderomotive force represents in fact the time average back-reaction of the high-frequency oscillations on the
low-frequency density inhomogeneities. The ponderomotive
force is the spatial gradient of a potential, i.e.
$\mathbf{f}=\nabla U$ with $U=|E|^{2}/8\pi n_{0}$,
and this force can be shown to act primarily on the electrons. In addition to the separation
of time-scales inherent to the Zakaharov equations, one can further exploit a separation of length-scales
by considering the particular case of non-local three-wave interactions involving
an ion-sound mode having a wave-vector and frequency given by $[\mathbf{q},\Omega(\mathbf{q})]$
and two Langmuir modes, $[\mathbf{k},\omega(\mathbf{k})]$ and $[\mathbf{k}+\delta\mathbf{k},\omega(\mathbf{k}+\delta\mathbf{k})]$,
under the restriction $\delta k\ll k$ for non-local mode coupling.
Therefore, the two resonance conditions for mode coupling are $\mathbf{q}+\mathbf{k}=\mathbf{k}+\delta\mathbf{k}$
and $\Omega(\mathbf{q})+\omega(\mathbf{k})=\omega(\mathbf{k}+\delta\mathbf{k})$.
By Taylor expanding the second condition, we obtain that
$\Omega(\mathbf{q})= \delta \mathbf{k}.\nabla_{k}\omega(\mathbf{k})$ and using the first one,
which is $\mathbf{q}=\delta \mathbf{k}$,
the resonance conditions for the non-local
three-wave interactions are found to reduce to only one condition $\Omega(\mathbf{q}) =\mathbf{q}.\mathbf{v}_{g}$, i.e.
$\pm qc_{s}=3 \lambda_{De}^{2}\omega_{pe}\mathbf{q}.\mathbf{k}$, which is
similar to a wave-particle resonance condition $\omega(\mathbf{k}) =\mathbf{k}.\mathbf{v}$.
Non-local resonant interactions between the large-scale and the small-scale
waves (quasi-particles) are therefore expected to mediate local diffusive energy transfer among the small-scale waves
in analogy with wave-particle interactions. By local diffusive energy transfer we mean that the spread of spectral
energy associated with the small-scale waves involves small steps in wave-number space, having $\delta k\ll k$ and that this spread is a diffusion.
Since the first Zakharov equation has the form of a Schrodinger equation describing the dynamics of Langmuir waves
in the a slowly varying potential $n/n_{0}$, it is standard to apply WKB analysis to it. In order to do that
it is convenient to use a generalization of the Fourier transform to spatially inhomogenous systems
\citep{2009JMP....50a3527G}. The
window transform of $E(\mathbf{x},t)$ is defined by
\begin{equation}
\Gamma[E(\mathbf{x})]\equiv\widehat{E}(\mathbf{x},\mathbf{k})=\int d\mathbf{x}_{0}w(\epsilon^{*}|\mathbf{x}-\mathbf{x_{0}}|)E(\mathbf{x}_{0})e^{i\mathbf{k}.\mathbf{x}_{0}},
\end{equation}
which is a kind of wavelet transform that in the case where the window function $w(x)=\exp(-x^{2})$ is called a Gabor transform. The parameter $1/\epsilon^{*}$ is the width of the window which is taken to be much smaller than the characteristic length of the inhomogeneity but
much larger than the wavelength of the Langmuir waves that propagate in the inhomogeneous medium.
Note that the Gabor transform can be viewed as a localized Fourier
 transform around $x$ with support $1/\epsilon^{*}$ : when $\epsilon^{*}$ goes to zero, the filtering width goes to infinity and the Gabor transform becomes a Fourier transform.
Applying this transform to Equation (\ref{z1}), we have
\begin{equation}\label{gb}
\frac{i}{\omega_{pe}}\frac{\partial \widehat{E}_{\mathbf{k},\mathbf{x}}}{\partial t}=
\left[\frac{3}{2}\lambda_{De}^{2}k^{2}+\left(\frac{n}{2n_{0}}\right)-\nabla\left(\frac{n}{2n_{0}}\right)\right]
\widehat{E}_{\mathbf{k},\mathbf{x}}+i \nabla\left(\frac{n}{2n_{0}}\right).\frac{\partial {\widehat{E}_{\mathbf{k},\mathbf{x}}}}
{\partial \mathbf{k}} -3\lambda_{De}^{2}i\mathbf{k}.\nabla\widehat{E}_{\mathbf{k},\mathbf{x}}
\end{equation}
Multiplying by $\widehat{E}^{*}_{\mathbf{k},\mathbf{x}}$, the imaginary part of (\ref{gb}) gives
the Liouville equation for the small-scale wave energy density.
Information on the phases of the small-scale waves is now lost as these waves are treated as
 quasi-particles.
Moreover, let us introduce the wave-action density, defined as
\begin{equation}
N_{\mathbf{k}}=\frac{W_{\mathbf{k},\mathbf{x}}}{\omega(\mathbf{k})},
\end{equation}
and the group velocity $\mathbf{v}_{g}=3\lambda_{De}^{2}\omega_{pe}\mathbf{k}$, therefore,
the WKB limit of the Zakharov equations
are written as in \citet{1967PlPh....9..719V}:
\begin{equation}\label{ty1}
\frac{\partial N_\mathbf{k}}{\partial t}+\mathbf{v}_{g}.\frac{\partial N_{\mathbf{k}}}{\partial \mathbf{x}}
-\frac{1}{2}\omega_{pe}\nabla \left(\frac{n}{n_{0}}\right).\frac{\partial N_{\mathbf{k}}}{\partial \mathbf{k}}=0,
\end{equation}

\begin{equation}
n_{0}m_{i}\left[\frac{\partial ^{2}}{\partial t^{2}}-c_{s}^2\nabla^{2}\right]\left(\frac{n}{n_{0}}\right)
=\frac{1}{2\omega_{pe}}\nabla^{2}\int d\mathbf{k}N_{\mathbf{k}}.
\end{equation}
The \citet{1972JETP...35..908Z} equation were first written by \citet{1967PlPh....9..719V}
in this WKB form, which corresponds to restricting the
consideration to non-local three-wave coupling with $k_{L},k_{L'}\sim k \gg k_{S}=q$ in
the process $L+S=L'$.
This system is similar to the field/particle system of Eqs.(1)-(2) we started from, so that quasilinear diffusion equations can be derived from it.
Linearizing the first equation (\ref{ty1}) with
respect to $N_{\mathbf{k}}=N_{\mathbf{k}}^{0}+\delta N_{\mathbf{k}}$,
$i(\Omega(\mathbf{q})-\mathbf{q}.\mathbf{v}_{g})\delta N_\mathbf{k}=\left(\omega_{pe}n/2n_{0}\right)\mathbf{q}.\partial N^{0}_{\mathbf{k}}/\partial \mathbf{k}$,
and substituting $\delta N_\mathbf{k}$ in the second equation,
one obtains the dispersion relation connecting the frequency $\Omega(\mathbf{q})$ and the wave-vector
$\mathbf{q}$ of the ion-acoustic wave which now includes an imaginary part $\gamma_{\mathbf{q}}$.
By considering the imaginary part $\gamma_\mathbf{q}$ of $\Omega(\mathbf{q})$ to be small compared
with the frequency $\Omega(\mathbf{q})$, i.e. $\gamma_\mathbf{q}\ll \Omega(\mathbf{q})$ then
\begin{equation}\label{mod}
\gamma _{\mathbf{q}}=\frac{\pi \omega_{pe}^{2} q}{8n_{0}m_{i}c_{s}}\int d\mathbf{k}\, \mathbf{q}
.\frac{\partial N^{0}_{\mathbf{k}}}{\partial \mathbf{k}}
\delta (\Omega(\mathbf{q})-\mathbf{q}.\mathbf{v}_{g}),
\end{equation}
where in the last expression
$\Omega(\mathbf{q}) = \pm qc_{cs}$ is used as dispersion relation. The imaginary part $\gamma_{\mathbf{q}}$ is the coefficient describing the Landau damping or growth of the ion-acoustic waves on the Langmuir
plasmons. When it is positive the coefficient
(\ref{mod}) is the growth rate of the so-called modulational instability \citep{1978PhR....43...43T}.

Moreover, we have seen above that Langmuir plasmons undergo a diffusion in wavenumber space
in the presence of a spectrum of density fluctuations.
 Here, this diffusion process in wave-number space occurs
  as a result of the action of the stochastic refraction force of the sound modes on the quasi-particles.
 The diffusion coefficient is given by $
D^{*}_{ij}=(\pi \omega_{pe}^{2}/4)\int d\mathbf{q} q_{i}q_{j}S_{n}(\mathbf{q})\delta (\Omega(\mathbf{q})-\mathbf{q}.\mathbf{v}_{g})$.
 Therefore, in the non-local limit and under the random phase approximation, the Zakharov equations take the common
 quasilinear diffusive form given by
\begin{equation}\label{h1}
\frac{\partial N(\mathbf{k})}{\partial t}=\frac{\partial }{\partial k_{i}}D^{*}_{ij}\frac{\partial N(\mathbf{k})}{\partial k_{j}},
\end{equation}
\begin{equation}\label{h2}
\frac{\partial S_{n}(\mathbf{q})}{\partial t}=\gamma_{\mathbf{q}}S_{n}(\mathbf{q}).
\end{equation}
On the other hand, we have seen that wave-number diffusion is a source of broadening
of the wave-particle resonance.
Therefore, the quasilinear diffusion equations describing the interactions between Langmuir waves and electrons
will involves a resonance function $R(\omega-\mathbf{k}.\mathbf{v}, \Delta \omega)$ which is broadened
due to the mode-coupling
of Langmuir waves with sound waves described by Eqs.(\ref{h1})-(\ref{h2}).
As shown in Section 4, the wave-particle resonance width is given by
$\Delta\omega=(D^{*}_{ij}v_{i}v_{j})^{1/3}$.

To summarize, we have obtained the following resonance-broadened model of weak-turbulence:
\begin{equation}\label{p1}
\frac{\partial f(\mathbf{v})}{\partial t}= \frac{\partial }{\partial v_{i}}
D_{ij} \frac{\partial f(\mathbf{v})}{\partial v_{j}},
\end{equation}
 \begin{equation}\label{p2}
\frac{\partial N(\mathbf{k})}{\partial t}=\gamma_{\mathbf{k}}N(\mathbf{k}),
\end{equation}
\begin{equation}\label{p3}
\frac{\partial S_{n}(\mathbf{q})}{\partial t}=\gamma_{\mathbf{q}}S_{n}(\mathbf{q}),
\end{equation}
where
\begin{equation}\label{q1}
D_{ij}=\frac{4\pi e^{2}\omega_{pe}}{m_{e}^{2}}\int d\mathbf{k} \frac{k_{i}k_{j}}{k^{2}}N(\mathbf{k})R(\omega_{pe}
-\mathbf{k}.\mathbf{v},\Delta \omega),
\end{equation}
\begin{equation}\label{q2}
\gamma_{\mathbf{k}}=\omega_{pe}^{3}\int d\mathbf{v}
\frac{\mathbf{k}}{k^{2}}.\frac{\partial f(\mathbf{v})}{\partial \mathbf{v}} R(\omega_{pe}-\mathbf{k}.\mathbf{v},\Delta \omega),
\end{equation}
\begin{equation}\label{q3}
\gamma _{\mathbf{q}}=\frac{\pi \omega_{pe}^{2} q}{8n_{0}m_{i}c_{s}}\int d\mathbf{k}\, \mathbf{q}.
\frac{\partial N_{\mathbf{k}}}{\partial \mathbf{k}}
\delta (\Omega(\mathbf{q})-\mathbf{q}.\mathbf{v}_{g}).
\end{equation}
 and where the
 wave-particle resonance width entering $R(\omega_{pe}-\mathbf{k}.\mathbf{v},\Delta \omega)$ is given by
\begin{equation}\label{yu}
\Delta\omega=(D^{*}_{ij}v_{i}v_{j})^{1/3},
\end{equation}
which is related to the dynamical level of sound waves fluctuations through
\begin{equation}\label{yz}
D^{*}_{ij}=\frac{\pi \omega_{pe}^{2}}{4}\int d\mathbf{q}\, q_{i}q_{j}S_{n}(\mathbf{q})\delta (\Omega(\mathbf{q})-\mathbf{q}.\mathbf{v}_{g}),
\end{equation}
where $\Omega(\mathbf{q})=\pm qc_{s}$.
In this formulation of the \cite{1967PlPh....9..719V} equations, the effect of mode-coupling between the Langmuir plasmons and the sound-waves is taken into account through a broadening of the wave-particle resonances.

Notice that the level of sound waves $S_{n}(\mathbf{q})$ entering (\ref{yz}), which is also
involved in the expression for the broadening width (\ref{yu}),
is depending on time via Eq.(\ref{p3}) which involves $\gamma_{\mathbf{q}}$, and hence $\partial N_{\mathbf{k}}/\partial \mathbf{k}$ .
This dynamical broadening of the wave-particle interaction
reflects the self-consistent level of Langmuir wave scattering by an evolving
spectrum of sound modes in the beam-plasma system. Disregarding time evolution of the sound waves
given by Eq.(\ref{p3}) results in Eqs.(\ref{p1})-(\ref{p2}), together with Eqs.(\ref{q1})-(\ref{q2})
and with Eqs.(\ref{yu})-(\ref{yz}), being a description of the relaxation of an electron beam
in a \emph{prescribed} level of density fluctuations unaffected by the Langmuir waves.
These density fluctuations may not necessarily be
produced by sound waves for an arbitrary dispersion relation $\Omega(\mathbf{q})$.
These density fluctuations may not necessarily
be produced by waves when no such dispersion relation exists, in which case
we formally need to take $
D^{*}_{ij}=\pi \omega_{pe}^{2}/4\int d\mathbf{q}\, \int d\Omega q_{i}q_{j}S_{n}(\mathbf{q},\Omega)\delta (\Omega-\mathbf{q}.\mathbf{v}_{g})$
instead of (\ref{yz}) in the expression for the wave-particle resonance width ($\ref{yu}$).

\section{Summary and discussion}\label{sec:sum}
In general, the resonance function of quasilinear theory has a finite width and can be
written in the form given by
\begin{equation}\label{ref3}
R(\omega-kv,\Delta\omega)=\int_0^\infty dt \langle e^{ikx-i\omega t}\rangle,
\end{equation}
where $\Delta \omega$ is the resonance width.
The quasilinear diffusion coefficient then takes the form
\begin{equation}\label{dkk}
D=\frac{e^{2}}{m_{e}^{2}} \int \int dk \, d\omega  \,
S_{E}(k,\omega) \, R(\omega-kv,\Delta \omega),
\end{equation}
where $S_{E}(k,\omega)$ is the spectrum of electric field fluctuations.
We are interested in electric field fluctuations which are produced by Langmuir waves
whose dispersion relation is $\omega(k)\approx \omega_{pe}=\sqrt{4\pi n_{e}e^{2}/m_{e}}$,
and hence we can write $S_{E}(k,\omega) = S_{E}(k) \, \delta (\omega-\omega_{pe})$.
The spectral energy density of Langmuir waves is given by $
W(k)=2(S_{E}(k)/8\pi)$, where the factor of two accounts for the electric energy and the
 kinetic energy of the thermal electrons participating in the wave
motion. Hence, the quasilinear diffusion coefficient is given by
\begin{equation}\label{diff2}
D= \frac{4\pi
e^{2}}{m_{e}^{2}} \int dk
W(k)R(\omega_{pe}-kv,\Delta \omega) \,\,\, .
\end{equation}
By conservation laws, statistical acceleration of the particles by the waves, or their deceleration, means
damping or growth of the waves. Therefore, the rate of Landau damping or growth of Langmuir waves is given by
\begin{equation}\label{lgd2}
\gamma= \frac{\omega_{pe}^{3}}{2k^{2}} \int dv
k \frac{\partial f}{\partial v}R(\omega_{pe}-kv,\Delta
\omega).
\end{equation}
The quasilinear diffusion equations, involving a broad resonance function, have
the standard form given by the system
\begin{equation}\label{eqdiffuse2}
\frac{\partial f(v)}{\partial t}=\frac{\partial }{\partial v} D \frac{\partial
f(v)}{\partial v} \,\,\, ,
\end{equation}
\begin{equation}\label{eqdiffuse3}
\frac{\partial W(k)}{\partial t}= 2 \gamma W(k),
\end{equation}
which by construction respects conservation of the energy.
In practice, it is useful to approximate the broad resonance
function, entering the quasilinear diffusion equations, by a square function
\begin{eqnarray}
R(\omega_{pe}-kv,\Delta \omega) =\left\{\begin{array}{ll}
              \frac{\pi}{\Delta\omega}, & \qquad |\omega_{pe}-kv|<\Delta \omega, \\
              0, & \qquad |\omega_{pe}-kv|>\Delta \omega,
              \end{array}
\right.
\label{eq:square}
 \end{eqnarray}
having the property that $
R(\omega_{pe}-kv,\Delta \omega)\buildrel {\Delta \omega \rightarrow 0} \over
{\longrightarrow} \pi \delta (\omega_{pe}-kv)$, the sharp resonance being obtained from $x=vt$
and $\omega=\omega_{pe}$ in (\ref{ref3}) .

Turbulent scattering of particles results in velocity fluctuations $\Delta v$ around the free-streaming
trajectory so that we have instead $x=(v+\Delta v)t$ in (\ref{ref3}). When these fluctuations evolve diffusively, $\langle\Delta v^{2}\rangle=2Dt$, then
the width of the resonance is given by
\begin{equation}\label{dup2}
\Delta \omega=(Dk^{2})^{1/3}.
\end{equation}
In \cite{1966PhFl....9.1773D} work $D$ is the quasilinear diffusion coefficient itself, so that $\Delta \omega$ is
a function of the electric field amplitude.
Pitch-angle scattering off turbulent magnetic field
fluctuations can also broaden the wave-particle resonance leading to
$
\Delta \omega=(2E_{c}D^{(\mu)}k^{2}/m_{e})^{1/3} $
which tantamount to replacing $D\rightarrow(2E_{c}/m_{e})D^{(\mu)}$
in (\ref{dup2}), with $D^{(\mu)}$ the pitch-angle diffusion coefficient.
Resonance broadening due to collisional pitch-angle scattering, $D^{(\mu)}$ taking its collisional
value, is discussed by \cite{1992wapl.book.....S}.
Langmuir wave scattering
off background density fluctuations can also result in broadening of the resonance.
In particular, when the beam-driven Langmuir wave undergoes stochastic refraction, its wave-number
evolves diffusively so that $k\rightarrow k+\Delta k$ in (\ref{ref3}). Assuming diffusive
evolution in wave-number space $\langle\Delta k^{2}\rangle=2D^{*}t$, the resonance
width is then given by
\begin{equation}
\Delta\omega=(D^{*}v^{2})^{1/3},
\end{equation}
where $D^{*}$ is the diffusion coefficient in wave-number space.
Finally, let us discuss the consequence of
resonance broadening on the development of the beam instability.
From the broadening frequency $\Delta \omega$, one can also define a scale $\Delta v$
in velocity-space given by
\begin{equation}\label{resv}
\Delta v=\frac{\Delta\omega}{k}.
\end{equation}
Using the square as a representation of the resonance function,
the growth rate of the beam-plasma instability can also be written in terms of $\Delta v$, as
\begin{equation}\label{marg}
\gamma (\Delta v)\approx \frac{\pi \omega_{pe}^{3}}{2k^{2}}\frac{f(\omega_{pe}/k +\Delta v/2)-f(\omega_{pe}/k-\Delta v/2)}{\Delta v}\equiv \frac{\pi \omega_{pe}^{3}}{2k^{2}}
\frac{\Delta f}{\Delta v}.
\end{equation}
In other words, the basic time-scale $\gamma$ for the development of the beam instability
is now a function of the resonance width $\Delta \omega$ through $\Delta v$.
Resonance broadening filters out
variations in the electron phase-space
distribution which are smaller than $\Delta v$. As phase-space derivatives
also yield the rate of growth of the beam instability in Eq.(\ref{marg}), $\gamma$ will
in turn depend on which scale the phase-space has been filtered by the underlying
broadening processes. Notice that when $\Delta v\rightarrow 0$, the expression for the growth
rate Eq.(\ref{marg}) including broadening reduces to the standard Landau damping/growth rate, i.e.
$\gamma(\Delta v\rightarrow 0)\rightarrow (\pi \omega_{pe}^{3}/2k^{2})\partial f/\partial v|_{v=\omega_{pe}/k}$.
This means that for a given distribution function
$f(v)$ including a beam of extent $\Delta v_{b}$, the growth rate $\gamma (\Delta v)$ can either be positive or negative
depending on the value of the filtering scale $\Delta v$ in velocity space as illustrated below.
\begin{figure}
\begin{center}
\includegraphics[width=0.5\textwidth]{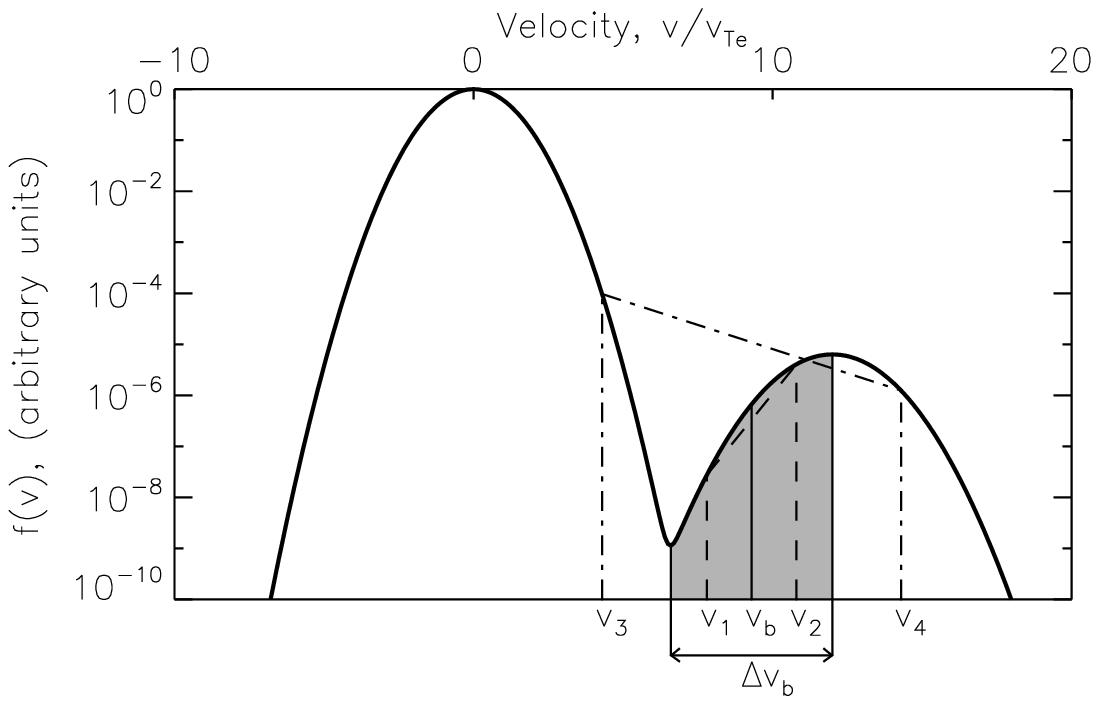}
\end{center}
\caption{
An example of weakening and
 possible suppression of the beam instability by resonance
 broadening. The electron distribution $f(v)$ is the sum of a thermal Maxwellian
of temperature $T_e$ and a beam with velocity $v_b=9 v_{Te}$ for $v_{Te}$ the electron thermal velocity. The shaded grey,
corresponds to the region of positive slope $\partial f/\partial v>0$ in the distribution
function. It is centered around $v_{b}$ with a width $\Delta v_{b}$.
For a resonance width such that $\Delta v= v_2-v_1 < \Delta v_b$,
the growth rate $\gamma(\Delta v)\propto \Delta f/\Delta v>0$
and, hence, the instability proceeds, but at a slower rate as the effective slope is smaller.
For an even larger resonance width,
covering $\Delta v=v_4-v_3 > \Delta v_b$, we have $\gamma(\Delta v)\propto \Delta f/\Delta v<0$ and, hence, the instability
is suppressed, i.e. the beam is stable to the generation of Langmuir waves.}
 \label{nico}
\end{figure}
Let us consider qualitatively the situation of the beam instability depicted in Fig 1.
Due to the presence of the electron beam, Langmuir waves are expected to
grow with a phase speed of the order of the beam speed $\sim v_{b}$. However, the presence of background scattering
has the effect on broadening the resonant
region in velocity space. For small resonance width $\Delta
v=v_{1}-v_{2}\ll \Delta v_{b}$, the average slope of electron phase-space
distribution is still positive,
therefore the particles loose energy and the waves grow but at a
reduced rate. The effective growth rate is given by Equation (\ref{marg}) for $\gamma(\Delta v)$.
For a large resonance width i.e. when $\Delta v=v_{3}-v_{4}\gg
\Delta v_{b}$, the average slope becomes positive and the waves
damp as now $\gamma(\Delta v)<0$. In the situation of Fig (1), the condition $
\Delta v \gg \Delta v_{b}$
sets the regime of the suppression of the beam instability due to
the presence of background scattering of the particles or of the waves. The
growth rate (\ref{marg}) turns out to be negative despite the
existence of a region of positive slope for the distribution
function, because small-scale features in velocity space, smaller than $\Delta v$, and
 which should act
as a reservoir of free-energy for the Langmuir waves to grow, have
been effectively filtered out.
Resonance broadening is a general mechanism at the origin of the
weakening and possible suppression of
the beam-plasma instability. As a consequence, it can be used as a tool to
explain stable inverted populations of fast electrons commonly observed in-situ
in the solar-wind (P. Kellogg, private communication). We refer to the work by \citet{1986ApJ...308..954L},
where electron distribution functions are observed not to exhibit the strong plateauing predicted by quasi-linear models.

With respect to Langmuir waves scattering off background density fluctuations,
we obtain the following criterium
for the suppression of the beam instability :
\begin{equation}
D^{*}>\frac{\omega_{pe}^{3}}{v_{b}^{2}}\left(\frac{\Delta v_{b}}{v_{b}}\right)^{3}
\end{equation}
When needed, a more accurate criterium can be obtained by directly solving the marginal
stability $\gamma(\Delta \omega)=0$ condition, Equation (\ref{marg}).

To summarize, we have shown that resonance broadening provide a unified framework
accounting for both the effects of particle and wave scattering during the quasilinear
relaxation of electron beams in turbulent plasmas. Resonance broadening is essentially equivalent
 to filtering-out small scale features of the electron phase-space distribution. As a beam
 is such a small scale feature in velocity space, resonance broadening provides
 an intuitive picture for the stabilization of the beam instability. Starting directly from the electrostatic Zakharov equations,
 we derived a resonance broadened version of weak-turbulence theory, where
 broadening of the wave-particle resonance occurs as a result of
 non-local mode coupling to ion-sound waves. In this case, broadening is non-linear and dynamical reflecting
 the self-consistent level of ion-sound waves generated
  by Langmuir waves in the system. In the future, we plan to extend this formalism to account
 for the mode-coupling of beam generated Langmuir waves to kinetic Alfven waves.


%
%
%
%
%
%
%

\begin{acknowledgments}
This work is  supported by a STFC
consolidated grant. Financial support by the European Commission through the "Radiosun"
(PEOPLE-2011-IRSES-295272) and HESPE (FP7-SPACE-2010-263086) is
gratefully acknowledged. N.H.B thanks Boris Breizman for his useful remarks regarding the
manuscript.
\end{acknowledgments}

%
%
%
%
%
%
%
%
%

\bibliographystyle{agufull08}
\end{article}
%
%
%
%
%
%

%
%


\end{document}